\documentclass[prl,aps,preprint,showpacs]{revtex4}

\usepackage{graphicx}

\begin{document}

\title{Possible manifestation of spin fluctuations in the temperature behavior of resistivity in
$Sm_{1.85}Ce_{0.15}CuO_4$ thin films}

\author{S. Sergeenkov$^{1,2}$,  A.J.C. Lanfredi$^1$ and F.M. Araujo-Moreira$^{1}$}

\affiliation{$^{1}$ Departamento de F\'{i}sica e Engenharia
F\'{i}sica, Grupo de Materiais e Dispositivos,\\ Centro
Multidisciplinar para o Desenvolvimento de Materiais
Cer\^amicos,\\ Universidade Federal
de S\~ao Carlos, S\~ao Carlos, SP, 13565-905 Brazil\\
$^{2}$ Bogoliubov Laboratory of Theoretical Physics, Joint
Institute for Nuclear Research,\\ Dubna 141980, Moscow Region,
Russia}

\date{\today; Accepted for publication in JETP Letters}

\begin{abstract}
A pronounced step-like (kink) behavior in the temperature
dependence of resistivity $\rho (T)$ is observed in the
optimally-doped $Sm_{1.85}Ce_{0.15}CuO_4$ thin films around
$T_{sf}=87K$ and attributed to manifestation of strong spin
fluctuations induced by $Sm^{3+}$ moments with the energy $\hbar
\omega _{sf}=k_BT_{sf}\simeq 7meV$. In addition to fluctuation
induced contribution $\rho _{sf}(T)$ due to thermal broadening
effects (of the width $\omega _{sf}$), the experimental data are
found to be well fitted accounting for residual (zero-temperature)
$\rho _{res}$, electron-phonon $\rho _{e-ph}(T)=AT$ and
electron-electron $\rho _{e-e}(T)=BT^2$ contributions. The best
fits produced $\omega _p=2.1meV$, $\tau _0^{-1}=9.5\times
10^{-14}s^{-1}$, $\lambda =1.2$, and $E_F=0.2eV$ for estimates of
the plasmon frequency, the impurity scattering rate,
electron-phonon coupling constant, and the Fermi energy,
respectively.
\end{abstract}


\maketitle

{\bf 1. Introduction.} Despite numerous investigations on many
different physical properties of electron-doped superconductors
(EDS), these interesting materials continue to attract attention
of both experimentalists and theoreticians alike, especially as
far as their low-temperature anomalies are concerned (see,
e.g.,~\cite{1,2,3,4,5,6,7} and further references therein). Of
particular interest is $Sm$-based EDS. Since $Sm$ has a larger ion
size than $Ce$, $Pr$ and $Nd$, it is expected that paramagnetic
scattering contribution to low-temperature behavior of
$Sm_{2-x}Ce_xCuO_4$ should be much stronger than in
$Pr_{2-x}Ce_xCuO_4$ and $Nd_{2-x}Ce_xCuO_4$. Recently~\cite{7}, by
using a high-sensitivity home-made mutual-inductance technique we
managed to extract with high accuracy the temperature profiles of
penetration depth $\lambda (T)$ in high-quality optimally-doped
$Sm_{1.85}Ce_{0.15}CuO_4$ (SCCO) thin films grown by the pulsed
laser deposition technique. We found that above and below
$T=0.22T_C$ our films are best-fitted by a linear~\cite{6} and
quadratic~\cite{2} dependencies, respectively, with physically
reasonable values of $d$-wave node gap parameter $\Delta
_0/k_BT_C=2.07$ and paramagnetic impurity scattering rate $\Gamma
/T_C=0.25(T_C/\Delta _0)^3$. We also noticed that the boundary
temperature ($T=0.22T_C$) which demarcates two scattering
mechanisms (pure and impure) lies very close to the temperature
where strong enhancement of diamagnetic screening in SCCO was
observed~\cite{4} and attributed to spin-freezing of $Cu$ spins.
Moreover, the above crossover temperature remarkably correlates
with the temperature where an unexpected change in the field
dependence of the electronic specific heat in PCCO crystals was
found~\cite{5} and attributed to the symmetry change from nodal to
gapped.

It should be mentioned also that in addition to their unusual
pairing properties, EDS exhibit some anomalous normal state
behavior far above $T_C$ with a noticeable presence of both
electron-phonon and electron-electron contributions~\cite{8,9,10}.
Recent inelastic neutron scattering experiments~\cite{11,12} on
low-energy spin dynamics (for the energy spectrum ranging from
$1meV$ to $10meV$) in $Pr_{0.88}LaCe_{0.12}CuO_{4-\delta}$ (PLCCO)
clearly demonstrated the evolution of PLCCO from
nonsuperconducting antiferromagnet (with the Neel temperature
$T_N=210K$) to optimally doped superconductor (with $T_C=24K$).
Besides, a step-like intensity increase was observed at about
$T_{sf}=80K$ and linked to the manifestation of low-energy ($\hbar
\omega _{sf}=k_BT_{sf}\simeq 6.5meV$) long-range antiferromagnetic
(AFM) spin fluctuations in the excitation spectrum induced by
$Pr^{3+}$ moments through $Cu^{2+}-Pr^{3+}$ interaction~\cite{13}.

\begin{figure}
\centerline{\includegraphics[width=8.0cm,angle=90]{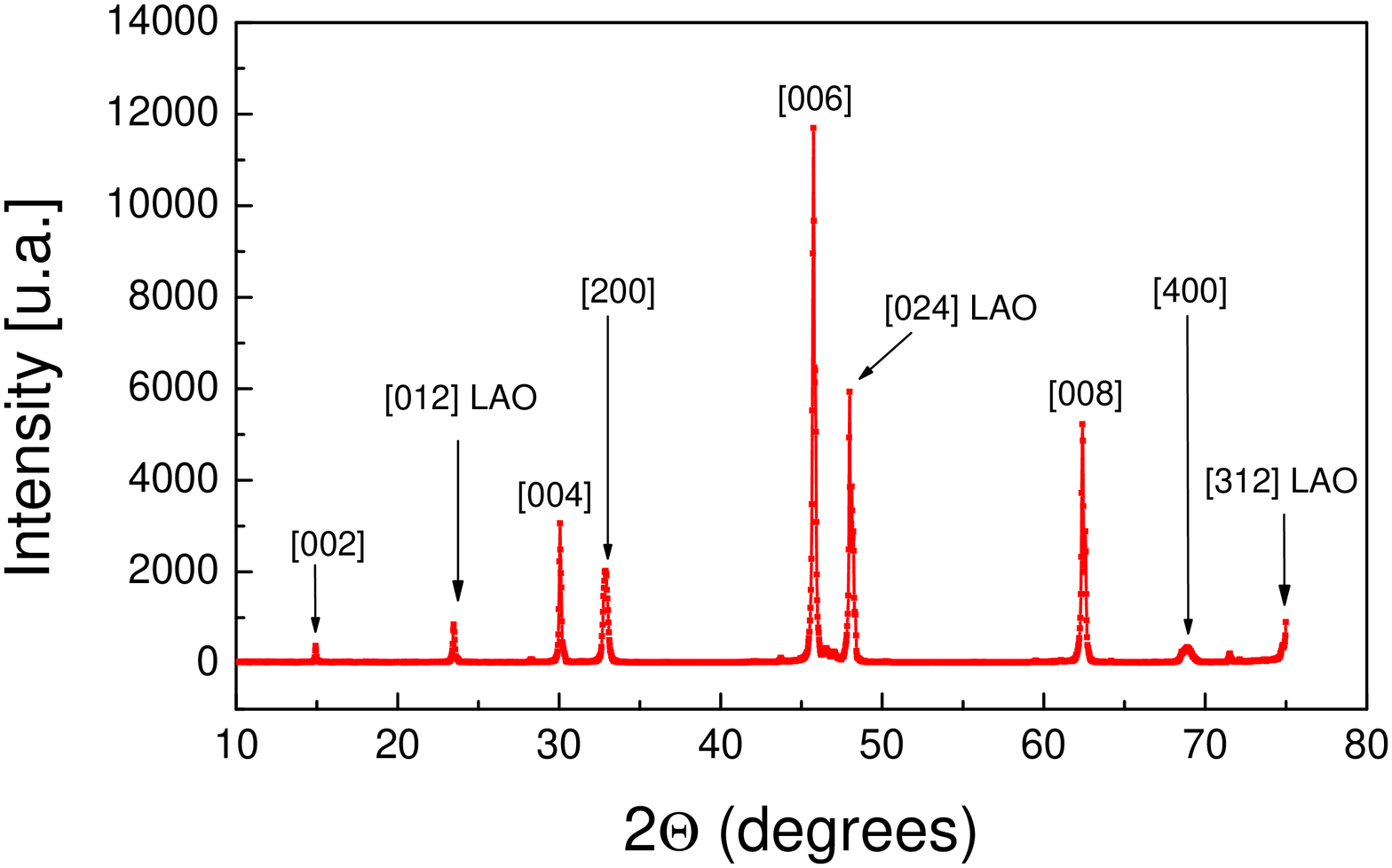}}
\vspace{0.5cm} \caption{X-ray diffraction spectrum of SCCO films.
} \label{fig:fig1}
\end{figure}

In this Letter we present our latest results on the temperature
behavior of resistivity $\rho (T)$ for the same optimally-doped
$Sm_{1.85}Ce_{0.15}CuO_4$ films~\cite{7}, paying special attention
to their normal state properties. In addition to the expected
contributions from the electron-phonon and electron-electron
scattering processes, we also observed an unusual kink like
behavior of $\rho (T)$ around $T=87K$ very similar to the one seen
in inelastic neutron scattering data~\cite{11,12}. Given that $Sm$
has a larger ion size than $Pr$ and assuming that the long-range
AFM correlations should be even stronger in thin films (than in
single crystals), we attribute the appearance of this kink in our
SCCO films to the manifestation of thermal excitations due to spin
fluctuations induced by $Sm^{3+}$ moments through
$Cu^{2+}-Sm^{3+}$ interaction.

{\bf 2. Results and Discussion.} A few SCCO thin films ($d=200nm$
thick) grown by pulsed laser deposition on standard $LaAlO_3$
substrates were used in our measurements (for more details on our
samples including their other physical properties, see Ref.7). All
samples showed similar and reproducible results. The structural
quality of the samples was verified through X-ray diffraction
(Fig.~\ref{fig:fig1}) and scanning electron microscopy together
with energy dispersive spectroscopy technique. To account for a
possible magnetic response from substrate, we measured several
stand alone pieces of the substrate. No tangible contribution due
to magnetic impurities was found. The electrical resistivity $\rho
(T)$ was measured using the conventional four-probe method. To
avoid Joule and Peltier effects, a dc current $I=1mA$ was injected
(as a one second pulse) successively on both sides of the sample.
The voltage drop $V$ across the sample was measured with high
accuracy by a $KT256$ nanovoltmeter.

\begin{figure}
\centerline{\includegraphics[width=8.0cm]{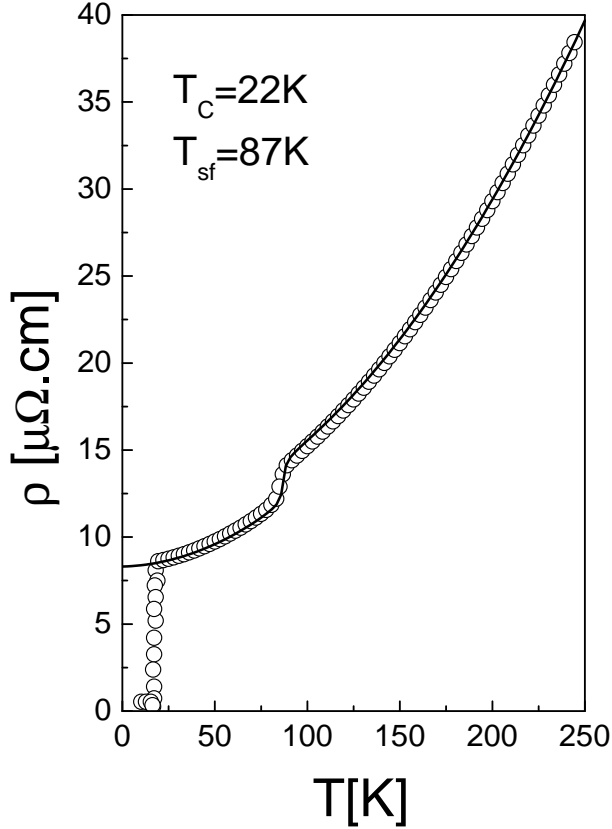}}\vspace{0.5cm}
\caption{Temperature dependence of the resistivity $\rho (T)$
measured for a typical SCCO thin film ($T_C =22K$). The solid line
is the best fit according to Eq.(3).} \label{fig:fig2}
\end{figure}

Fig.~\ref{fig:fig2} shows the typical results for the temperature
dependence of the resistivity $\rho (T)$ in our SCCO thin films.
Quite a pronounced step (kink) is clearly seen around $T=87K$.
Since, according to the X-ray diffraction spectrum
(Fig.~\ref{fig:fig1}), our films do not show any low-energy
structural anomalies, it is quite reasonable to assume that the
observed kink can be attributed to the manifestation of long-range
AFM spin fluctuations induced by $Sm^{3+}$ moment with the
characteristic energy $\hbar \omega _{sf}=7meV$ (corresponding to
an effective temperature $T_{sf}=\hbar \omega _{sf}/k_B=87K$).
More specifically, to account for fluctuation induced thermal
broadening effects (of the width $\omega _{sf}$) we suggest a
Drude-Lorentz type expression for this contribution (Cf. Ref.14):
\begin{equation}
\rho _{sf}(T)=\rho _{res}\int_{-\omega _{sf}}^{\Omega (T)
 -\omega _{sf}} \frac{\omega _{sf} d\omega}{\pi (\omega ^2+
 \omega _{sf}^2)}
=\rho _{res}\left [\frac{1}{4}+\frac{1}{\pi}\tan^{-1}\left
(\frac{T-T_{sf}}{T_{sf}}\right ) \right ]
\end{equation}
where $\rho _{res}$ is the residual contribution given by
\begin{equation}
\rho _{res}=\frac{1}{\omega _p^2\epsilon _0\tau _0}
\end{equation}
with $\omega _p$ being the plasmon frequency, $1/\tau _0$ the
corresponding scattering rate, and $\epsilon _0=8.85\times
10^{-12} F/m$ the vacuum permittivity. Notice that $\rho
_{sf}(0)=0$.

The temperature dependence in Eq.(1) comes from the cutoff
frequency $\Omega (T)=U(T)/\hbar$ which accounts for spin
fluctuations with an average thermal energy
$U(T)=\frac{1}{2}C<u^2> \simeq k_BT$ where~\cite{15} $C$ is the
force constant of a two-dimensional harmonic oscillator, and
$<u^2>$ is the mean square displacement of the magnetic $Sm$ atoms
from their equilibrium positions.

After trying many different temperature dependencies, we found
that our SCCO films are rather well fitted (solid line in
Fig.~\ref{fig:fig2}) using the following expression for the
observed resistivity:
\begin{equation}
\rho (T)=\rho _{res}+\rho _{sf}(T)+\rho _{e-ph}(T)+\rho _{e-e}(T)
\end{equation}
where the other two terms in the rhs of Eq.(3) are related to the
electron-phonon contribution~\cite{8} $\rho _{e-ph}(T)=AT$ with
\begin{equation}
A=\frac{\lambda k_B}{\hbar \omega _p^2\epsilon _0}
\end{equation}
and to the electron-electron contribution~\cite{9,10} $\rho
_{e-e}(T)=BT^{2}$ with
\begin{equation}
B=\frac{k_B^2}{\hbar \omega _p^2\epsilon _0E_F}
\end{equation}
Here, $\lambda $ is the electron-phonon coupling constant, and
$E_F$ the Fermi energy.

Using the experimentally found values of $\rho _{res}=8.8\mu
\Omega cm$, $A=0.14\mu \Omega cm/K$, $B=0.0012\mu \Omega cm/K^2$,
and $T_{sf}=87K$, the best fits through the data points produced
$\omega _p=2.1meV$, $\tau _0^{-1}=9.5\times 10^{-14}s^{-1}$,
$\lambda =1.2$, and $E_F=0.2eV$ for very reasonable~\cite{8,9,10}
estimates of the plasmon frequency, the impurity scattering rate,
electron-phonon coupling constant, and the Fermi energy,
respectively.

In summary, a pronounced step-like (kink) behavior in the
temperature dependence of resistivity $\rho (T)$ was observed in
the optimally-doped $Sm_{1.85}Ce_{0.15}CuO_4$ thin films around
$T=87K$ and attributed to manifestation of strong spin
fluctuations resulting in thermally activated displacement of $Sm$
atoms. The normal state experimental data were successfully fitted
by accounting for the residual, fluctuation, electron-phonon and
electron-electron contributions.

%\section{acknowledgments}

We gratefully acknowledge financial support from Brazilian
agencies FAPESP and CNPq.


\newpage

\begin{thebibliography}{99}

\bibitem{1} N.P. Armitage, D.H. Lu, D.L. Feng, C. Kim, A. Damascelli, K.M. Shen, F. Ronning,
Z.-X. Shen, Y. Onose, Y. Taguchi, and Y. Tokura, Phys. Rev. Lett.
{\bf 86}, 1126 (2001).
\bibitem{2} A. Snezhko, R. Prozorov, D.D. Lawrie, R.W. Giannetta, J. Gauthier, J. Renaud, and  P. Fournier,
 Phys. Rev. Lett. {\bf 92}, 157005 (2004).
\bibitem{3} W. Yu, B. Liang and R.L. Greene, Phys. Rev. B {\bf 72}, 212512 (2005).
\bibitem{4} R. Prozorov, D.D. Lawrie, I. Hetel, P. Fournier, and R.W. Giannetta, Phys. Rev. Lett. {\bf 93}, 147001 (2004).
\bibitem{5} Hamza Balci and R.L. Greene, Phys. Rev. Lett. {\bf 93}, 067001 (2004).
\bibitem{6} A. Zimmers, R.P.S.M. Lobo, N. Bontemps, C.C. Homes, M.C. Barr, Y. Dagan, and R.L. Greene,
Phys. Rev. B {\bf 70}, 132502 (2004).

\bibitem{7} A.J.C. Lanfredi, S. Sergeenkov, and F.M. Araujo-Moreira,
Phys. Lett. A {\bf 359}, 696 (2006).

\bibitem{8} J.A. Skinta, M.-S. Kim, T.R. Lemberger, T. Greibe, and
M. Naito, Phys. Rev. Lett. {\bf 88},  207005 (2002).

\bibitem{9} Y. Dagan, M.M. Qazilbash, C.P. Hill, V.N. Kulkarni, and R.L. Greene, Phys. Rev. Lett.
{\bf 92},  167001 (2004).

\bibitem{10} Dinesh Varshney, K.K. Choudhary, and R.K. Singh, J.
Supercond. {\bf 15}, 281 (2002).

\bibitem{11} S.D. Wilson, Shiliang Li, Pengcheng Dai, Wei Bao, Jae-Ho Chung, H.J. Kang, Seung-Hun Lee,
Seiki Komiya, Yoichi Ando, and Qimiao Si, Phys. Rev. B {\bf 74},
144514 (2006).

\bibitem{12} E.M. Motoyama, G. Yu, I.M. Vishik, O.P. Vajk, P.K. Mang, and M.
Greven, Nature {\bf 445}, 186 (2007).

\bibitem{13} A.N. Lavrov, H.J. Kang, Y. Kurita, T. Suzuki, S. Komiya, J.W.
Lynn, S.-H. Lee, Pengcheng Dai, and Y. Ando, Phys. Rev. Lett. {\bf
92}, 227003 (2004).

\bibitem{14} C.C. Homes, R.P.S.M. Lobo, P.Fournier, A. Zimmers, and
R.L. Greene, Phys. Rev. B {\bf 74}, 214515 (2006).

\bibitem{15} Ch. Kittel, {\it Introduction to Solid State Physics} (John Wiley and Sons, New
York, 1996), p. 632.

\end{thebibliography}
\end{document}